\documentclass[%
  preprint,
superscriptaddress,
 amsmath,amssymb,
aps,
pra,
floatfix,
]{revtex4-2}

\usepackage{graphicx}
\usepackage{subfigure}
\usepackage{dcolumn}
\usepackage{bm, bbm}
\usepackage{xcolor}
\usepackage{physics}
\usepackage{comment}

\newcommand{\bR}{{\bm R}}

\newcommand{\bP}{{\bm P}}

\newcommand{\bp}{{\bm p}}

\newcommand{\bd}{{\bm d}}
\newcommand{\bPi}{{\bm \Pi}}

\newcommand{\hGamma}{{\hat{\bm{\Gamma}}}}

\newcommand{\edit}{}

\begin{document}

\title{Symmetry Breaking as Predicted by a Phase Space Hamiltonian with a Spin Coriolis Potential}

\author{Nadine C. Bradbury}
\affiliation{Department of Chemistry, Princeton University, Princeton, NJ 08544 USA}
\author{Titouan Duston}
\affiliation{Department of Chemistry, Princeton University, Princeton, NJ 08544 USA}
\author{Zhen Tao}
\affiliation{Department of Chemistry, Princeton University, Princeton, NJ 08544 USA}
\author{Jonathan I. Rawlinson}
\affiliation{Department of Mathematics, Nottingham Trent University, Nottingham, United Kingdom}
\author{Robert Littlejohn}
\affiliation{Department of Physics, University of California, Berkeley, California 94720, USA}
\author{Joseph Subotnik} 
\email{subotnik@princeton.edu}
\affiliation{Department of Chemistry, Princeton University, Princeton, NJ 08544 USA}

\date{\today}

\begin{abstract}
We perform electronic structure calculations for a set of molecules with degenerate spin-dependent ground states  ($^3$CH$_2$, $^2$CH$_3^{\bullet}$, $^3$O$_2$) going beyond the Born-Oppenheimer approximation and accounting for nuclear motion. According to a phase space (PS) approach that parametrizes electronic states ($\ket{\Phi}$) and electronic energies ($E$) by nuclear position {\em and} momentum (i.e., $\ket{\Phi(\bR,\bP)}$ and $E(\bR,\bP)$), we find that the presence of degenerate spin degrees of freedom  leads to broken symmetry ground states. More precisely, rather than a single degenerate minimum at $(\bR,\bP) = (\bR_{min}, 0)$, the ground state energy has two minima at $(\bR,\bP) = (\bR_{min}',\pm \bP_{min})$ (where $\bR_{min}'$ is close to $\bR_{min}$), dramatically contradicting the notion that the total energy of the system can be written in separable form as $E = \frac{\bP^2}{2M} + V_{el}$.  Although we find that the broken symmetry solutions have small barriers between them for the small molecules, we hypothesize that the barriers should be macroscopically large for metallic solids, thus offering up a new phase-space potential energy surface for simulating the Einstein-de Haas effect.
\end{abstract}

\maketitle

\section{Introduction \label{sec:intro}}
Most modern schemes for efficiently propagating mixed quantum-classical nuclear-electronic dynamics do not conserve the total (nuclear + electronic + spin) angular momentum -- at least not in a meaningful sense.   For instance, within traditional classical Born-Oppenheimer (BO) dynamics, if one studies a system with an even number of electrons and a time-reversal invariant  nondegenerate ground state, it is straightforward to show that one conserves the {\em nuclear} linear and angular momentum.   {\edit More precisely, although BO theory formally conserves the total momentum -- representing the total momentum by the nuclear momentum in the BO representation \cite{littlejohn:2023:jcp:angmom} -- for a system with an even number of electrons and a time reversal invariant  nondegenerate ground state, the electronic momentum is formally set to zero (and so momentum conservation is not really meaningful).  In this sense,  even if one incorporates some  beyond BO physics by including the Berry force (whenever the latter is nonzero) \cite{Culpitt2024_magnetic_berry_force, Gross2014_nonBO} and indirectly accounts for how the electronic momentum interacts with the nuclear momentum (leading to a conservation law \cite{xuezhi:2023:total_ang_bomd, Gross2022_exact_factor,helgaker:2022:jcp_conservation_laws_magnetic_field}), 
one still cannot use BO theory to accurately capture momentum exchange and conservation in most cases. }
This state of affairs becomes even worse if we consider angular momentum, where we must account for angular momentum transfer between nuclear, electronic orbital, and electronic {\em spin} degrees of freedom. 
{\edit Moreover, in the presence of spin, } if one wishes to go beyond single-state dynamics,
the most standard nonadiabatic approach, Tully's original surface hopping algorithm (TSH)\cite{Tully1990}, becomes ill-defined for a doublet in the presence of  spin-orbit coupling (SOC);  in such a case, not only are the adiabatic states ill-defined because of  Kramers' degeneracy, but one finds complex-valued derivative couplings, and therefore complex nuclear momentum upon momentum rescaling--which requires a new understanding of momentum in surface hopping simulations\cite{bian:2022:pssh,yanzewu:2024:jcp:pssh_conserve}.


The difficulties cited above present major problems for \textit{ab initio} modeling of some of the most interesting and scientifically pressing spin dynamics related phenomena. After all,  spin-phonon coupling determines the superconductivity transition temperature in most heavy-fermion materials,\cite{Pellegrini2024,Carbotte1990}  spin-phonon coupling also defines the spin-lifetime (e.g. usefulness) of spin-qubits based on diamond nitrogen-vacancy centers,\cite{MacQuarrie2013,Nakane2024} spin-phonon and electron-phonon coupling in ferromagnetic nickel lead to highly non-thermal distributions and energy transfer during non-equilibrium dynamics,\cite{Beaurepaire1996,Maldonado2020} and spin-phonon coupling is one putative mechanism behind the origin of the chirality induced spin selectivity (CISS)  effect.\cite{Bloom2024, Mtangi2017} In order to understand the systems above, atomistic modeling is essential.  That being said, quite often, the electronic structure of systems at a single geometry is often hard enough (without considering dynamics)\cite{Herrmann2018_spintronics}.  Nonetheless, for realistic simulations (especially with disorder), there is no alternative to explicitly modeling the coupling of nuclear, electronic and spin degrees of freedom, a dire need that necessitates the development of a computationally feasible theory that goes beyond the BO framework. 

Let us now directly address the physics that is not captured by BO dynamics. 
As early as 1929, it was recognized that, in order to recover the correct vibrational  spectrum of small molecules, 
one would need to account for how nuclear motion perturbs the electronic spin.\cite{VanVleck1929} 
The resulting spin-rotation coupling \cite{VanVleck1951, Curl1965,Howard1970,Willatt2024} is distinct from spin-orbit coupling, and has been observed spectroscopically and applied to a variety of  small molecules,  including O$_2$\cite{Kayama1967} and CH$_3$,\cite{nesbit:1997:spinrot} -- which are treated in this paper.  The form of the perturbation is:
\begin{eqnarray}
    H_{SR} = -a_0\frac{ g \alpha^2}{2m_e} \sum_A\sum_i \frac{Z_A}{|\hat{\bm{r}}_i-\bm{R}_A|^3} ((\hat{\bm{r}_i}-\bm{R}_A)\times(\frac{1}{M_A}\bm{P}_A))\cdot\hat{\bm{s}}_i \label{eq:HSR}
\end{eqnarray} 
Here, $a_0$ is the Bohr radius and $\alpha = e/hc \approx 1/137$ is the fine structure constant, and {\edit  $g\approx 2.0023$ is the Land\'e $g$-factor of the electron.}  This perturbation is almost universally added to the spin-orbit coupling terms
\begin{eqnarray}
        H^{(1)}_{\mathrm{SO}} &=& a_0  \frac{g\alpha^2}{4m_e^2}\sum_A\sum_i \frac{Z_A}{|\hat{\bm{r}}_i-\bm{R}_A|^3}((\hat{\bm{r}}_i -\bm{R}_A)\times \hat{\bm{p}}_i)\cdot \hat{\bm{s}}_i \label{eq:Hso1} \\
        H^{(2)}_{\mathrm{SO}} &=& -a_0 \frac{g\alpha^2}{4 m_e^2} \sum_{ij} \frac{1}{|\hat{\bm{r}}_i-\hat{\bm{r}}_j|^3}((\hat{\bm{r}}_i -\hat{\bm{r}}_j) \times \hat{\bm{p}}_i)\cdot\hat{s_i} \label{eq:Hso2} 
\end{eqnarray}
and the electronic spin-other-orbit coupling term:
\begin{eqnarray}            
        H^{(2)}_{\mathrm{SOO}} &=& a_0 \frac{g\alpha^2}{2m_e^2} \sum_{ij} \frac{1}{|\hat{\bm{r}}_i-\hat{\bm{r}}_j|^3}((\hat{\bm{r}}_i -\hat{\bm{r}}_j)\times \hat{\bm{p}}_j)\cdot\hat{\bm{s}_i} \label{eq:Hso3} 
\end{eqnarray}

Both Eq. \ref{eq:HSR} and \ref{eq:Hso3} are ``spin-other-orbit coupling'' terms that assess how the {\edit spin of the target particle precesses in the presence the magnetic field produced by another source particle in the source particles's rest frame}; in Eq. \ref{eq:HSR}, the ``other'' particle is a nucleus; in Eq. 
\ref{eq:Hso3}, the other particle is an electron. 
Eqs.  \ref{eq:Hso1} and \ref{eq:Hso2} arise from investigating the magnetic field that the target spin particle feels in its own rest frame, as caused by the relative motion of a nucleus or electron   ---  noting that the factor of $\frac{1}{2}$ in both of Eq. \ref{eq:Hso1} and \ref{eq:Hso2} come from a  Thomas procession correction.   

Now, it must be noted that, if one seeks to separate the electronic Schrodinger equation from the nuclear Schrodinger equation in the spirit of  Born-Oppenheimer theory, there are other problems besides the presence of nuclear momentum in Eq. \ref{eq:HSR}  above. Most importantly, if one seeks the most accurate description of coupled nuclear-electronic dynamics, one must account for the  fact that because nuclei are never truly frozen, the electronic Schrodinger equation is being solved in a moving (non-inertial) frame, new forces necessarily arise--in particular a Coriolis term. Now, importantly there are actually {\em two} so-called Coriolis forces that arise: $(i)$ an orbital Coriolis force that arises from solving the motion of electrons in a moving frame and $(ii)$ a spin Coriolis force that arise from solving the electronic spin in a moving nuclear frame.  The final Coriolis potential is well known in the magnetic field and neutron  communities and takes the form (for a diatomic):\cite{Wick1948,Werner1979,Mashhoon1988,Moody1986, Wilson1975,Hehl1990,
 Shen2003, Matsuo2011, Geilhufe2022}
\begin{eqnarray}
    H_{\mathrm{SR}} =\omega\cdot\hat{\bm{j}} = \bm{\omega} \cdot (\hat{\bm{l}} + \hat{\bm{s}}),
\end{eqnarray}
for an angular nuclear velocity $\omega = \frac{|\bm{P}|}{M|\bm{R}|}$.
Often these terms are considered together, and physicists have attempted to estimate these terms from the Berry curvature within spin-unrestricted calculations.\cite{Steiner1994} 
That being said, there have also been quite a few recent papers in the literature examining the effect of specifically spin Coriolis forces on electronic dynamics, with an eye towards understanding chiral phonons. \cite{Juraschek2020,Geilhufe2023,Funato2024} 

At present, however, if one wishes to go beyond diatomics and/or systems with very small unit cells, there is no obviously tractable means to model the above physics for large system sizes, accounting fully the for the exchange of momentum between nuclei and electrons and spin. Moreover, as mentioned above, BO dynamics cannot be cured by including  Berry pseudo-magnetic force\cite{Culpitt2024_magnetic_berry_force, Gross2014_nonBO}, insofar as one  corrects only the nuclear but not the electronic observables, and the method is difficult to apply with degenerate states (which is essential for problems with spin states).  Perhaps the most successful  approach so far has been exact factorization\cite{Gross2022_exact_factor}, but  the latter cannot yet been applied to large \textit{ab initio} systems. An alternative technique is Ehrenfest dynamics and work with multiple electronic states\cite{doltsinis:2002:review}; although the common Ehrenfest equations of motion do not perfectly conserve momentum\cite{truhlar:2020:project_rot_nac} due to a truncation to a finite number of BO surfaces, momentum conservation can be recovered by  including the non-adiabatic Berry curvature\cite{coraline:2024:jcp:ehrenfest_conserve,Takatsuka2005,Krishna2007}. Nevertheless, while mean-field dynamics do capture some elements of Berry force\cite{xuezhi:2023:total_ang_bomd}, they too do not offer complete electronic feedback and mean-field dynamics cannot address thermodynamics averages (because Ehrenfest dynamics do not satisfy detailed balance\cite{Parandekar2005, Schmidt2008}). 
All of the problems above motivate an alternative, phase-space approach to electronic structure theory.

\subsection{Notation} A word is now appropriate about notation. Above, below and throughout this work, we will use bold font for all vector quantities, upper case letters for nuclear variables, such as $\bm{P}$, and hats for all electronic operators such as $\hat{\bm{p}}$. Indices $A,B \dots$ are used for nuclei and $i,j\dots$ are used for electrons.

\subsection{An Electronic Phase Space Hamiltonian }

The essence of a phase space approach is to avoid the pitfalls of BO and Ehrenfest theory by parameterizing the electronic Hamiltonian by both nuclear position ($\bR$) and nuclear momentum $(\bP)$ in a manner that conserves linear and angular momentum. In practice, one of the first electronic phase space Hamiltonians was developed by Shenvi\cite{Shenvi2009}, who was inspired by Berry's work on superadiabats.\cite{Berry1990_superadiabats}  Shenvi posited that, if one were prepared to work semiclassically with an electronic Hamiltonian that was parametrically dependent on both the nuclear positions $\{\bm{R}_A\}$ ( as in traditional Born-Oppenheimer theory) {\em and} on the nuclear momentum $\{ \bm{P}_A\}$, then the natural choice was to use the standard BO expansion {\edit but without the nuclear quantum operators}:
\begin{equation}
    \hat{H}_\mathrm{Shenvi}(\bm{R},\bm{P}) = \sum_{A, ijk} \frac{1}{2M_A}(\bm{P}_A\delta_{ij} - \bm{d}^A_{ij}) \cdot (\bm{P}_A\delta_{jk} - \bm{d}^A_{jk})\ket{\Phi_i}\bra{\Phi_k} + \sum_i E_i(\{\bm{R}\}) \ket{\Phi_i}\bra{\Phi_i}. \label{eq:Shenvi}
\end{equation}
Here,  $\hat{H}_{el}\ket{\Phi_i} = E_i\ket{\Phi_i}$ is the standard diagonalization of the electronic Hamiltonian, and $\bm{d}_{ij}^A = \bra{\Phi_i}\frac{\partial}{\partial \bm{R}_a}\ket{\Phi_j}$ is the derivative coupling vector (sometimes called the non-abelian gauge potential or non-abelian berry connection in physics\cite{Wilczek1984_gauge}).

Notably, in the absence of any coupling between spin and spatial degrees of freedom,  the derivative coupling satisfies four constraints\cite{littlejohn:2023:jcp:angmom,littlejohn:2024:jcp:moyal,yanzewu:2024:jcp:pssh_conserve}:
\begin{eqnarray}
 -i\hbar\sum_{A} {\bm d}^A_{jk} + \bra{\Phi_j}\hat{\bm{p}} \ket{\Phi_k} &=& 0,\label{eq:cond1}  \\
    \sum_A \bm \nabla_{A} {\bm d}^B_{jk}  &=& 0 
    \label{eq:cond2}\\
    -i\hbar\sum_{A}{\bm R}_{A} \times {\bm d}^A_{jk} + \bra{\Phi_j} \hat{\bm{l}} \ket{\Phi_k} &=& 0,\label{eq:cond3}\\
     -\sum_A \left({\bm R}_A \times \bm \nabla_{A} \bm{d}^{B\beta}_{jk}\right)_{\alpha} &
     =&  \sum_{\gamma} \epsilon_{\alpha \beta \gamma} \bm{d}^{B \gamma}_{jk},\label{eq:cond4} 
\end{eqnarray}
Physically, for $j \ne k$ (where one can apply the Hellman-Feynmann theorem), one can clearly see that 
Eqs. \ref{eq:cond1} and \ref{eq:cond3} arise from the translation invariance and rotational isotropy of the Hamiltonian, i.e. the fact that $[\hat{H}_{el},\hat{\bp} -i \hbar \sum_A \bm{\nabla}_A] = 0$ and $[\hat{H}_{el},\hat{\bm{l}} -i \hbar \sum_A \bR_A \times \bm{\nabla}_A ] = 0$. As a result, 
 the  adiabatic electronic states (that diagonalize the BO electronic Hamiltonian) can be expressed in coordinates relative to the nuclei: when the nuclei translate, the electronic eigenstates must be translated as well; when the nuclei rotate, the electronic eigenstates must be rotated as well. More generally,  for  all $j,k$ (including the diagonal case), Eqs. \ref{eq:cond1} and \ref{eq:cond3} are phase conventions for the adiabatic electronic states as a function of nuclear position. 
Eqs. \ref{eq:cond2} and \ref{eq:cond4} signify that, if we displace or reorient the entire molecule, the derivative couplings either are invariant (under translations) or transform correctly (under rotations). For many more details about these relations, see Ref. \citenum{littlejohn:2024:jcp:moyal}.

When performing simulations with Eq. \ref{eq:Shenvi}, it is clear that
one introduces nonzero electronic momentum $\hat{\bm{p}}$ (on account of Eq. \ref{eq:cond1}) and  one  introduces an electronic angular momentum $\hat{\bm{l}}$ (on account of Eq. \ref{eq:cond3}) {\edit into the resulting solution wavefunctions}. For this reason, one can show\cite{yanzewu:2024:jcp:pssh_conserve} that Shenvi-inspired dynamics go beyond Born-Oppenheimer dynamics and introduce electronic momentum transfer in a fashion whereby the total momentum is conserved. That being said, using the derivative coupling comes with several problems: the derivative coupling vector is not well defined in the case of degenerate states, will diverge at curve crossings during dynamics, and, frankly, is too expensive compute for all electron pairs and nuclei for realistic systems.

For all of the reasons above, we have recently proposed that one replace Eq. \ref{eq:Shenvi} with a simpler (more efficient and more robust)  equation whereby the most essential parts of the derivative coupling are distilled into a single one-electron operator, $\hat{\bm{\Gamma}}_A$. \cite{Tao2024_abinitio,Qiu2024_ERF,Tao2024_basisfree} In this case, the total Hamiltonian is written as 
\begin{equation}
    \hat{H}_{\mathrm{PS}}(\bm{R},\bm{P}) = \sum_{A}\frac{1}{2M_A}(\bm{P}_A - i\hbar\hat{\bm{\Gamma}}_A(\{\bm{R}\}))^2 + \hat{H}_{el}(\{\bm{R}\}). \label{eq:H_PS}
\end{equation}If the $\hat{\bm{\Gamma}}$ operator satisfies the analogues of Eqs. \ref{eq:cond1} - \ref{eq:cond4}, 
\begin{eqnarray}
    -i\hbar\sum_{A}\hat{\bm{\Gamma}}_{A} + \hat{\bm{p}} &=& 0,\label{eq:condG1}  \\
    \Big[-i\hbar\sum_{B}\frac{\partial}{\partial\bm{R}_B} + \hat{\bm{p}}, \hat{\bm{\Gamma}}_A\Big] &=& 0,\label{eq:condG2}\\
    -i\hbar\sum_{A}{\bm R}_{A} \times \hat{\bm \Gamma}_{A} + \hat{\bm{l}} &=& 0,\label{eq:condG3}\\
     \Big[-i\hbar\sum_{B}\left(\bm{R}_B \times\frac{\partial}{\partial \bm{R}_B}\right)_{\gamma} + \hat{\bm{l}}_{\gamma}, \hat{\bm{\Gamma}}_{A \delta}\Big] &=& i\hbar \sum_{\alpha} \epsilon_{\alpha \gamma \delta} \hat{\bm{\Gamma}}_{A \alpha},\label{eq:condG4}
\end{eqnarray} 
one can show that the resulting equations  still conserve total momentum under dynamics. {\edit Note that, in Eqs. \ref{eq:condG1}-\ref{eq:condG4}, we have expressed the $\hat{\bm{\Gamma}}$ operator in the language of second quantization,  where $\hat{\bp} \equiv \sum_i \hat{\bp_i}$ is a sum over all electrons, and product operators are similarly one-electron operators, e.g.  $\hat{\bm{l}}_{\gamma}\hat{\bm{\Gamma}}_{A \delta} = \sum_i \bm{l}_{\gamma}(\bm{\hat r}_i) 
\bm{\Gamma}_{A \delta}(\bm{\hat r}_i)$.
}

As derived in a recent previous work, \cite{Tao2024_basisfree}, the form of this $\hat{\bm{\Gamma}} = \hat{\bm{\Gamma}}^{'} +\hat{\bm{\Gamma}}^{"}$ operator can be split into two terms, one for total (nuclear + electronic) linear momentum, and one for total angular momentum. 
The first term ($\hat{\bm{\Gamma}}'$) form a set of electron translation factors (ETFs) that captures the drag experienced by an electron when a nucleus moves;  one would pick up such a term by transforming to a local center of mass frame (CM). While not as simple as translations, the  second term ($\hat{\bm{\Gamma}}"$) is a set of electronic rotation factors (ERFs) that can be similarly be thought of as a transformation to the body frame of a molecular rotation. 

Altogether, the final proposed form of $\Gamma$ is:
\begin{eqnarray}
    \hat{\bm{\Gamma}}'_A &=& \frac{-i}{2\hbar}(\theta_A(\hat{\bm{r}})\hat{\bm{p}} +\hat{\bm{p}}\theta_A(\hat{\bm{r}})) \label{eq:Gamma'}\\ 
    \hat{\bm{\Gamma}}^{''}_A &=&
    \sum_B \zeta_{AB}(\bm{R}_A - \bm{R}_B^0)\times (\bm K^{-1}_B\hat{\bm{J}}_B)
    \label{eq:Gamma''}\\
    \hat{\bm{J}}_B &=& \hat{\bm{J}}_B^{(l)} = \frac{-i}{2\hbar}((\hat{\bm{r}}-\bm{R}_B)\times(\theta_B(\hat{\bm{r}})\hat{\bm{p}}) + (\theta_B(\hat{\bm{r}})\hat{\bm{p}})\times(\hat{\bm{r}}-\bm{R}_B) )  \label{eq:JB_noS}
    \\
    \bm{R}_B^0 &=& \frac{\sum_{A}\zeta_{AB}\bm{R}_B}{\sum_{A}\zeta_{AB}} \label{eq:R_B^0}\\
    \bm K_B &=& \sum_A\zeta_{AB}(\bm{R}_A\bm{R}_A^{T}-\bm{R}_B^0\bm{R}_B^{0T} - (\bm{R}_A\bm{R}_A^T - \bm{R}_B^{0T}\bm{R}_B^0)\mathcal{I}_3), \label{eq:K_B}
\end{eqnarray}
where we have introduced two {\edit new} `locality factors,' {\edit in} an electronic density partitioning
\begin{equation}
    \theta_A(\hat{\bm{r}}) = \frac{Z_A e^{-(\hat{\bm{r}}-\bm{R}_A)^2/\sigma^2}}{\sum_B Z_B e^{-(\hat{\bm{r}}-\bm{R}_B)^2/\sigma^2}}, \label{eq:theta}
\end{equation}
and a nuclear locality term, $\zeta_{AB} = M_A e^{-(\bm{R}_A-\bm{R}_B)^2/8\sigma^2}$. The parameter $\sigma$ controls the hardness of the locality filter, about 2.7 au. The tensor $\bm K_B$ can be thought to be a atomically partitioned moment of inertia, where in Eq. \ref{eq:K_B} superscript $T$ refers to a transpose, and $\mathcal{I}_3$ is a identity matrix of dimension $3$. $\hat{\bm r}$ is an electronic operator for position in 3D Cartesian space. Note that in Eq. \ref{eq:JB_noS}, we introduce the notation $\hat{\bm{J}}_B^{(l)} $ to emphasize its definition based on the electronic orbital angular momentum. See Eq. \ref{eq:JB_S} below for a similar definition based on electronic spin. 

Two points are now worth emphasizing. First, note that, when we remove locality by setting $\zeta_{AB}$ to $M_A$ (i.e. setting $\sigma \to \infty$), $\bm K_B = \bm K_\mathrm{CM}$ becomes the traditional moment of inertia tensor; in effect, we have  transformed the traditional electronic Hamiltonian to the nuclear center of mass frame. Furthermore, in this same nonlocal limit, if we  assert that $Z_A \propto M_A $, then 
the nuclear kinetic energy operator for the phase space Hamiltonian takes the form,
\begin{eqnarray}
 \sum_{A}\frac{1}{2M_A}(\bm{P}_A - i\hbar\hat{\bm{\Gamma}}_A(\{\bm{R}\}))^2 &=& \sum_A \frac{|\bm{P}_A|^2}{2 M_{tot}} - 
  \frac{1}{M_{tot}}\sum_A  \bm{P}_A \cdot \hat{\bm{p}} + \frac{|\hat{\bm{p}}|^2}{2M_{tot}} \label{eq:lim_Coriolis}\\
    &-&\sum_A \frac{1}{M_A}((\bm{R}_A - \bm{R}_{\rm{CM}})\times \bm{P}_A)\cdot \bm K_{\rm{CM}}^{-1} \cdot(\hat{\bm{l}}_{\rm{CM}}) \nonumber \\
    &+&\frac{1}{2}\hat{\bm{l}}_{\rm{CM}} \cdot \bm K_{\rm{CM}}^{-1}\cdot\hat{\bm{l}}_{\rm{CM}}\nonumber 
\end{eqnarray}

Above, the interpretation is very interesting and can be understood by realizing that $\bP_A$ now captures the total (not just nuclear) momentum. Thus, the first three terms represent the kinetic energy of the nuclear center of mass plus the kinetic energy of the nuclei relative to the center of mass (and we have taken care to remove the electronic contribution here as we must).  The fourth and fifth terms  are the electronic Coriolis and centrifugal potentials that arise by working in the rotating, non-inertial nuclear frame. 

Second, it is important to emphasize that ignoring locality is fundamentally incorrect. For instance, if one considers two rigid bodies separated by large distance, it is nonsensical to consider their joint motion along one center of mass; for such a scenario, there should be two centers of mass.  Moreover, vibrational circular dichroism (VCD) spectroscopy{\edit \cite{Nafie_1992_NVP,Nafie_1983_CA,patchkovskii:2012:jcp:electronic_current,Takatsuka_2021_fluxconserve,Duston2024_vcd}} demonstrates that a single molecule undergoing internal motion (that does not change the center of mass) does exhibit a nonzero electronic momentum. However, in the limit of complete nonlocality from Eq. \ref{eq:lim_Coriolis} above, no such electronic momentum can be found.
In essence, the locality factors, $\theta$ and $\zeta$ in Eqs. \ref{eq:Gamma'} and \ref{eq:Gamma''} above aim to capture both the motion of the center of mass and some of the relative motion observed in each atom's individual body frame. Note that, with meaningful parameters, 
phase-space electronic structure theory captures excellent VCD spectra\cite{Duston2024_vcd,Tao:2024:jcp_vcd_current}.

\subsection{Incorporation of a Spin-Dependent Momentum Coupling (a.k.a. a Spin-Coriolis Force) }

All of the theory above has been published in a recent set of papers\cite{Qiu2024_ERF,Tao2024_basisfree,Bian2025_vib, Bian2024_inertial} and sets the stage for the present manuscript. Note that, in Refs. \cite{Tao2024_abinitio,Qiu2024_ERF,Tao2024_basisfree,Bian2025_vib,Bian2024_inertial}, we have not yet addressed the question of electronic spin. That being said, when a molecule {\edit with strong enough spin-orbit coupling} rotates, one fully expects that the resulting spin-adiabatic states should rotate with the molecule. This intuition implies that one must replace the $\Gamma$ analogs of Eqs. \ref{eq:condG3} and \ref{eq:condG4} with  Eqs. \ref{eq:cond3s} and \ref{eq:cond4s}:
\begin{eqnarray}
    -i\hbar \sum_A \bm{R}_A \times \hat{\bm{\Gamma}}_A + \hat{\bm{l}} + \hat{\bm{s}} &=& 0 \label{eq:cond3s} \\ 
    \left[ -i\hbar \sum_B \left( \bm{R}_B \times\frac{\partial}{\partial \bm{R}_B}\right)_\gamma + \hat{\bm{l}}_\gamma + \hat{\bm{s}}_\gamma, \,\, \hat{\bm{\Gamma}}_{A\delta}\right] &=& i\hbar \sum_\alpha \epsilon_{\alpha \gamma \delta} \hat{\bm{\Gamma}}_{A\alpha} \label{eq:cond4s}
\end{eqnarray}
For our purposes,  
the introduction of electronic spin necessitates that we change the resulting ERFs and phase space electronic Hamiltonian and replace Eq. \ref{eq:JB_noS} with
\begin{eqnarray}
        \hat{\bm{J}}_B &=& \frac{-i}{2\hbar}\left((\hat{\bm{r}}-\bm{R}_B)\times(\theta_B(\hat{\bm{r}})\hat{\bm{p}}) + (\theta_B(\hat{\bm{r}})\hat{\bm{p}})\times(\hat{\bm{r}}-\bm{R}_B) + 2\theta_B(\hat{\bm{r}})\hat{\bm{s}}\right) .\label{eq:JB_tot} \\
         &\equiv & \hat{\bm{J}}_B^{(l)} + \hat{\bm{J}}_B^{(s)} \\
        \hat{\bm{J}}_B^{(s)} &=& -\frac{i}{\hbar}\theta_B(\hat{\bm{r}})\hat{\bm{s}}.\label{eq:JB_S} 
\end{eqnarray}
Physically,  this term arises when one seeks to align the spin of a molecule with the body frame (which is not inertial).
As a result of aligning both the electronic coordinates and the spin to the body frame, one can show that, in the limit $\sigma \to \infty$, one recovers the correct Coriolis potential in Eq. \ref{eq:lim_Coriolis} where we replace $\hat{\bm{l}}_{\mathrm{CM}}$ with $\hat{\bm{l}}_{\mathrm{CM}} + \hat{\bm{s}}$. For this reason, henceforward, letting
\begin{equation}
    \hat{\bm{\Gamma}}_A''^{(s)} \equiv
    \sum_B \zeta_{AB}(\bm{R}_A - \bm{R}_B^0)\times (\bm K^{-1}_B\hat{\bm{J}}_B^{(s)}) \label{eq:Gamma_s}
\end{equation}
we will refer to $-i \hbar \bP \cdot \hat{\bm{\Gamma}}_A''^{(s)} /M$ as a spin-Coriolis potential. 
At this juncture, however, a crucial point emerges: note that there is no $\alpha^2$ prefactor in Eqs. \ref{eq:H_PS} and \ref{eq:Gamma_s} which must be compared with Eq. \ref{eq:HSR}.  In other words, by moving to a rotating frame, we  that the effective spin Coriolis potential {\em  appears to be 18,000 times larger than the raw spin-rotation coupling term!}

{\edit Admittedly, the inclusion of a spin-Coriolis term as a part of the total momentum-coupling in a phase space Hamiltonian hinges on strong enough spin-orbit coupling in the electronic subsystem, and  this assumption is not applicable to particularly light molecules (including some in this paper). That being said, the assumption of an intermediate spin-orbit coupling is valid for most large molecules and extended systems (see Section \ref{sub:validity} for further clarification).} 

At present, to our knowledge, no one has yet explored the consequences of the spin-component of $\hat{\bm{\Gamma}}^{''}$ in Eq. \ref{eq:Gamma_s} in any practical manner for realistic molecules or materials. During the preparation of this manuscript, Polkovnikov and co-workers submitted an article examining the consequences of such spin terms using a Shenvi-like Hamiltonian (Eq. \ref{eq:Shenvi}) rather than an approximate Hamiltonian (Eq. \ref{eq:H_PS}) for a model problem with spin\cite{Polkovnikov2014_geometric,Polkovnikov2025_MBOA} in a magnetic field (where problems of degeneracy do not arise).
Crucially, however,  one cannot meaningfully apply Eq. \ref{eq:Shenvi} in the absence of a magnetic field on account of degeneracy; in such a case, one must use an approximation, e.g. Eq. \ref{eq:H_PS}. Moreover, given that Eqs. \ref{eq:Gamma'}-\ref{eq:theta} already yield phase space potential energy surfaces (via Eq. \ref{eq:H_PS}) with observable consequences vis-a-vis VCD spectra \cite{Duston2024_vcd} and electronic momentum\cite{Tao2024_basisfree}, one must wonder what will be the consequences of using Eq. \ref{eq:JB_S} instead of  Eq. \ref{eq:JB_noS} so as to address the question of spin. 

Moreover, considerations of spin are obviously of critical importance for at least three reasons. First, spin is a crucial form of angular momentum that directly couples to external magnetic fields and plays a key role in magnetism (see more about the Einstein de Haas effect in Sec. \ref{sec:disc} below). Second, the phase space electronic Hamiltonian in Eq. \ref{eq:H_PS} differs from BO by a small term, and therefore one can expect to find the largest effects on systems with closely spaced BO energy levels (with an energy gap smaller than the phase space perturbation terms). Obviously, spin-systems represent a clear and obvious set of degenerate molecular states.  
Third, even for systems whose ground state is normally a singlet, whenever one encounters an avoided crossing of some sort, there is usually a low lying triplet and the temptation to unrestrict and mix singlet and triplet,\cite{Burton2019} a condition for which we know there is a dynamically meaningful Berry force.\cite{Bian2022_berry_unrestricted}  This transition from a restricted (singlet) ground state to an unrestricted (mixed singlet-triplet) ground state occurs at the so-called Coulson-Fischer point,\cite{CoulsonFischer1949} where we expect to find at least two low-lying  electronic states and a lot of static correlation. Again, in the presence   
of a phase space electronic Hamiltonian approach, one might expect to find new physical manifestations of electronic-nuclear correlation. 
With this background in mind, our goal in what follows is to consider a small set of degenerate molecular systems (doublets and triplets) and explore the resulting electronic systems as a function of both $\bR$ and $\bP$ within a phase space electronic Hamiltonian approach, focusing on the effects of the relevant spin terms. 

\section{Results}

\subsection{The Physical Interpretation of Broken Symmetry Stationary Points \label{sec:interpretation}}

Our focus below will be on the characterization of stationary points, i.e. energetic minima, of the phase space energy $E_{PS}(R,P)$. 
Mathematically, we must imagine diagonalizing the phase-space electronic Hamiltonian,

\begin{eqnarray}
    \hat{{H}}_{\mathrm{PS}} \psi = E_{PS} \psi
\end{eqnarray}
and evaluating the energy
\begin{eqnarray}
    E_{PS}= \langle\psi| \hat{{H}}_{\mathrm{PS}} |\psi\rangle
    \label{eq:Eav}
\end{eqnarray}
Minima of this expectation value clearly satisfy:
\begin{eqnarray}
    \frac{\partial E_{PS}}{\partial \bm{R}} = 0 \label{eq:dEdR}\\
    \frac{\partial E_{PS}}{\partial \bm{P}} = 0 \label{eq:dEdP_blank}
\end{eqnarray}

While Eq \ref{eq:dEdR} above is the standard condition for finding an energetic minimum, Eq. \ref{eq:dEdP_blank} is not as common. In particular, for a separable classical Hamiltonian, where one can write ${H} = \bP^2/2M + V$, this equation is obviously satisfied only for $\bP = 0$. That being said, for a non-separable phase space Hamiltonian, as we will show below, one can find that Eq. \ref{eq:dEdP_blank} is satisfied for $\bP = \bP_{min} \ne 0$, which constitute broken symmetry points.

To physically interpret such stationary points in phase space, note that, according to Hamilton's equations, $\dot{\bR} = \frac{\partial E_{PS}}{\partial \bP}$. Therefore, at a stationary point $(\bR_{min}, \bP_{min})$ satisfying Eq. \ref{eq:dEdP_blank}, the nuclei composing a molecule or material are \underline{\em not} moving. In this regard, one must recognize the crucial difference between the canonical momentum ($\bP$) and the kinetic momentum $(\bPi = \bP - i\hbar\langle \hGamma \rangle)$ that arises in the presence of a magnetic field; in our case, there is a magnetic field caused by the internal motion of the electrons. Note that a simple derivate of Eq. \ref{eq:Eav} shows that
\begin{eqnarray}
    \sum_A \frac{\partial E_{\mathrm{PS}}}{\partial \bm{P}_A} = \sum_A \frac{\bm{P}_A}{M_A} - \frac{i\hbar \langle \hat{\bm{\Gamma}}_A\rangle}{M_A} \label{eq:dEdP}
\end{eqnarray}
Therefore, at a stationary point, 
\begin{eqnarray}
\label{eq:PAmin}
\bP_{A}^{min} = i\hbar\langle\hGamma_A\rangle,
\end{eqnarray}
which is equivalent to $\bPi_{min} = 0$. 

Finally, note that even though the nuclei are not actually moving at a stationary point, according to Eqs. \ref{eq:PAmin}, \ref{eq:cond1} and \ref{eq:cond3s}, it follows that 
\begin{eqnarray}
    \langle \hat{\bm{p}}\rangle &=& i\hbar \sum_A \langle \hat{\bm{\Gamma}}_A \rangle  = \sum_A \bP_{A}^{ min}\label{eq:Gamma=p}\\
    \langle \hat{\bm{l}} + \hat{\bm{s}} \rangle&=& i\hbar \sum_A \langle \bm{R}_A \times \hat{\bm{\Gamma}}_A \rangle  = \sum_A \bR_A^{min} \times \bP_{A}^{ min} \label{eq:Gamma=l+s}
\end{eqnarray}
In other words, if  $\bP_{min} \ne 0$ at a stationary point, we must conclude that there is a nonzero {\em electronic} linear and angular momentum at equilibrium -- a set of affairs that is normally associated with superconductivity rather than normal quantum chemistry. 

{\edit \paragraph{Implementation Details:}}

With this understanding in mind, let us now investigate four molecules/molecular fragments, H$_2$,O$_2$, CH$_2$ and CH$_3$, focusing on the coordinates in phase space $(\bR,\bP)$ that  minimize the phase space energy in Eq \ref{eq:H_PS}. The phase space Hamiltonian in Eq. \ref{eq:H_PS} has been implemented in both PySCF\cite{pyscf} and Q-Chem,\cite{qchem} with the data below presented only using the $\bm{P}\cdot\hat{\bm \Gamma}$ Coriolis term and not including the $\hat{\bm\Gamma}^2$ centrifugal energy; the latter terms are  $\propto \hat{\bm{s}}^2$ and merely lead to in a global energy shifts. {\edit For all results below, we have  implemented the one-electron spin-orbit operator given by eq. \ref{eq:Hso1}, but for convenience, we have not implemented the relatively smaller contributions from the two-electron terms in Eqs. \ref{eq:Hso2}-\ref{eq:Hso3}. We report full configuration interaction (FCI) calculations within an atomic orbital basis as well as a few comparisons to generalized Hartree Fock (GHF) results
in order to make sure we avoid any anomalies arising from electron-electron correlation and an incomplete wavefunction \cite{helgakerbook}; as such, our results  directly address the electron-nuclear correlation problem that is treated by a PS Hamiltonian.}

\subsection{Diatomics: H$_2$ and O$_2$\label{sec:diatomics}}

\begin{figure*}
    \centering
    \includegraphics[width=0.8\linewidth]{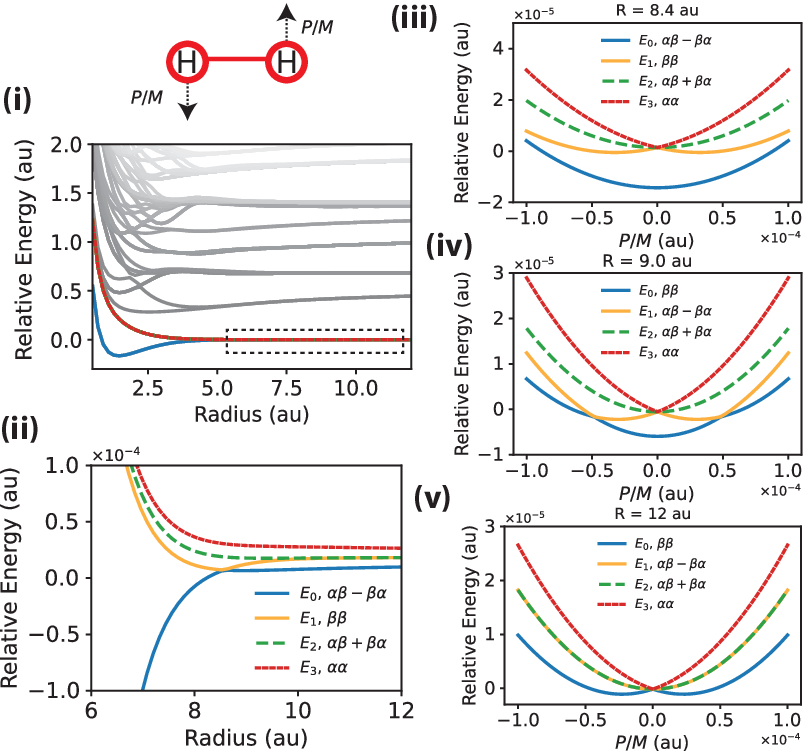}
    \caption{(i-ii) Relative energy of H$_2$ under the phase space Hamiltonian, $ E_\mathrm{H_2} ( \bm{R},\bm{P})-2E_H(\bP=0)$, using FCI and a cc-pVDZ basis. The lowest four states are labeled by their spin character at the optimal bond length. (iii-v) Relative energy dependence of the four lowest states  as a function of $P/M$ for different radii $R$, 8.4, 8.5 and 12 au. For subplots (iii)-(v), curves are labeled by their spin character at {\edit positive velocity,} $P/M  = 1\times 10^{-4}$au. {\edit Please note that the spin character of the $\alpha \alpha$ and $\beta \beta$ triplets will change signs at $P=0$.} }
    \label{fig:H2}
\end{figure*}

We begin our analysis with the simplest molecules, H$_2$ and O$_2$. Homonuclear diatomics have an obvious center of mass  that serves as the origin for angular momentum. Thus, we choose as our coordinates the radius $R$ of the molecule 
and the perpendicular velocity $\bP/M = \bm v$ that is consistent with said angular momentum (see Fig. \ref{fig:H2} for an illustration).  Conveniently, due to spatial inversion symmetry,  the overlap of  one spatial atomic orbital on the left hand nucleus A and  another spatial atomic orbital on the right hand nucleus B,  $\bra{\phi_A}\hat{\theta}_A\ket{\phi_B}$, is nearly independent of the locality factor $\sigma$. Moreover, as a practical matter, the matrix elements in this case are particularly simple to implement. { \edit Note that, for a diatomic molecule, both $\mathbf{L}_{CM}$ and $ \mathbf{K}_{CM}^{-1} \bm{v}$ are always perpendicular to the bond-direction (for any vector $\bm{v})$.
Thus, the natural extension of $H_{PS}$ given by Eq. \ref{eq:lim_Coriolis} to include spin is: $\sum_A \frac{1}{M_A}((\bm{R}_A - \bm{R}_{\rm{CM}})\times \bm{P}_A)\cdot \bm K_{\rm{CM}}^{-1} \cdot(\hat{\bm{l}}_{\rm{CM}})  \rightarrow \mathbf{L}_{CM} \cdot K_{\rm{CM}}^{-1} \cdot(\hat{\bm{l}}_{\rm{CM}} + \hat{\mathbf{s}})$ where only the components of the spin perpendicular to the bond appear in the dot product. For the explicit examples included below, let} the radius $R$ point along the x-axis, let the velocities along the y-axis, 
 $\ket{\phi_A \chi_A }$ and $\ket{\phi_B \chi_B }$ where $\phi$ and $\chi$ are spatial orbitals and spin orbitals respectively; in such a case, $\bra{\phi_A \chi_A } \bP \cdot \hGamma" \ket{\phi_B \chi_B } \propto  \frac{v}{R} \bra{\phi_A}\ket{\phi_B}
\bra{\chi_A }  \hat{\bm{s}}_z
\ket{\chi_B } = 0$, and thus only the energies of the $\alpha\alpha$ and $\beta\beta$ triplet configurations are affected by $\hGamma"$ and there is no diabatic coupling between the spin eigen-states of $\bm{\hat{s}}_z$ (in the absence of SOC).

\subsubsection{Bond dissociation in H$_2$}
Our first example is the most basic problem in electronic structure theory: the hydrogen molecule.  H$_2$ bond dissociation is a canonical multi-reference system taught to students, as the $S_0$ and three $T_1$ states become degenerate in energy at an infinite radius according to standard BO theory. Interestingly, however, on account of the  spin Coriolis term, this degeneracy is broken within phase space electronic structure theory. Instead, at long distances, the presence of a non-zero spin breaks the degeneracy leading to a ground state electronic structure with $\bP_{min} \ne 0$ and nonzero electronic angular momentum.

The description above is verified in Figure 1. In $(i)$, we show the overall potential energy surfaces for H$_2$ (with $v = 1\times 10^{-4}$ au) as a function of $R${\edit, the total internuclear distance}.  Next, in $(ii)$, we zoom in on the dissociation region beyond the Coulson-Fischer point  (beyond 2.28 au for cc-pVDZ) to see the triplet states splitting in energy according to $\langle m_s\rangle$. Note that, as $R$ grows larger and larger, there is indeed a crossover as one triplet state ($\beta\beta$) eventually becomes lower than the singlet state (due to the spin-Coriolis effect).
In $(iii-v)$, we further plot the phase space energy as a function of velocity at different  radii (8.4 au, 9.0 au and 12 au). 

\begin{figure*}
    \centering
    \includegraphics[width=\linewidth]{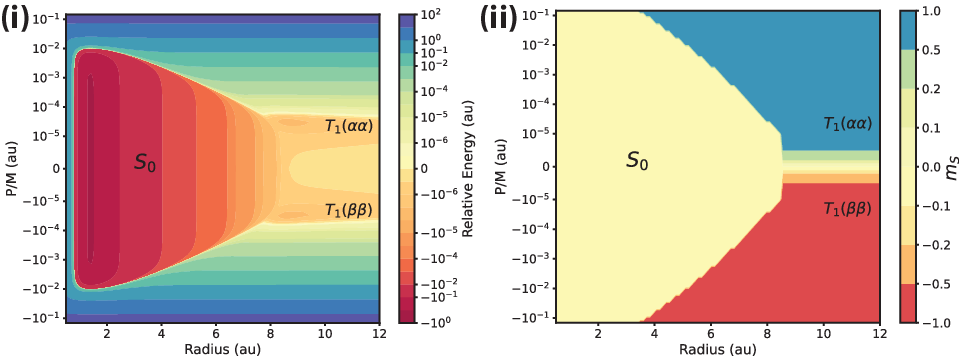}
    \caption{Heat maps  of the phase space energy of H$_2$ in an STO-3G basis as a function of bond distance a ($R$) and nuclear momentum over mass for the rotation ($P/M$).  See the schematic in Fig. \ref{fig:H2} above. In $(i)$, we plot the energy and we find three minima: the bonding minima (dotted contours), and long distance spin polarized minima due to rotating reference frame. The zero is taken as infinite separation with zero velocity, as in Figure \ref{fig:H2}.
    In $(ii)$, we identify the spin of the ground state.}
    \label{fig:contour}
\end{figure*}

All our intuition above is summed up in 
 Figure \ref{fig:contour}, where we draw heat maps of the relevant electronic structure quantities over phase space. In (i), we plot the relative  energies that arise from the phase space Hamiltonian in comparison to 2 isolated H atoms. Outside of the standard singlet bonding minima, we find two new distinct minima at long distances. At these two points in phase space, the total energy of the system is reduced by introducing a nonzero velocity and therefore an electronic angular momentum that couples to the spin. In $(ii)$, we plot the corresponding spin phase diagram, noting that the preferred spin state depends on both nuclear position and nuclear velocity.

\subsubsection{Triplet O$_2$}

\begin{figure*}
    \centering
    \includegraphics[width=\linewidth]{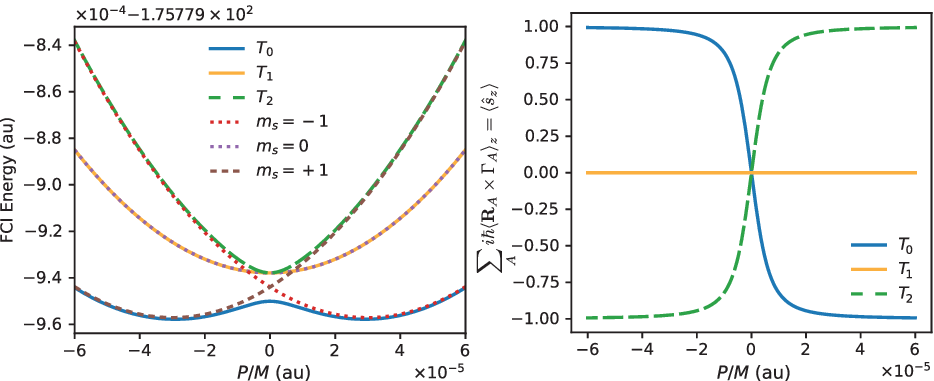}
    \caption{(i) FCI states in STO-3G basis for rotating O$_2$ groundstate $^3\Sigma_g$ (solids), and diabatized states separated by $\hat{\bm{s}}_z$ (dotted). (ii)  The expectation  value of the cross product of position with $\hat{\bm \Gamma}$ operator,  which is effectively a proxy for $\hat{\bm{s}}_z$.}
    \label{fig:O2}
\end{figure*}

Our second example is molecular oxygen, O$_2$, a prototypical example of a diatomic ground state triplet with reasonably strong SOC. 
For this system, we observe qualitatively the same spin-induced electronic structure  as we found molecular hydrogen. The primary change seen in Figure \ref{fig:O2} (versus Fig. \ref{fig:H2}(v)) is that at $v=0$ a zero field splitting due from the Bret-Pauli 1-electron spin orbit operator\cite{Neese2005_spin_orbit}, Eq \ref{eq:Hso1}. {\edit 
As noted above earlier in Sec. \ref{sec:diatomics}, for a diatomic molecule rotating in the $x-y$ plane,  only the $z-$component of the spin appears in the PS Hamiltonian; whether or not the $x-y$ component of the spin follows the molecular frame is not directly addressed by a PS approach.}
According to our phase space calculation, the zero field splitting at {\edit $P=0$} is roughly  the same size as the phase-space energy wells.  (Note though, that for O$_2$, the spin-orbit interaction is known to account for only two-thirds of the zero field splitting observed in EPR experiments, the remaining third resulting from the Bret-Pauli spin-spin dipole term\cite{Langhoff1977} -- which we have not included in our calculations; nor have we included 
the  Breit-Pauli 2-electron spin-orbit and spin-other-orbit coupling terms (Eqs. \ref{eq:Hso2} and \ref{eq:Hso3}).)

Finally in order to gain more intuition, in Fig. \ref{fig:O2} $(i)$, we plot the diabatic states as generated by diagonalizing the spin operator $\hat{\bm{s}_z}$ on top of the FCI adiabats, for the NIST standard bond length of 2.282 au\cite{Huber1979}. We note that the diabatic coupling, $\langle m_s = -1 | \hat{H}_\mathrm{PS}|m_s = +1\rangle$, is constant at all $\bP$. In (ii), we also plot the expected effect of $\hGamma$, via $\sum_A i\hbar\langle \bm{R}_A \times \hGamma_A\rangle_z = \langle \hat{\bm{s}}_z\rangle$. From these figures, we find a relatively smooth crossing between spin states, which is due to the reasonably large SOC. Were we to plot the same data for H$_2$ (with much less SOC and effectively zero field splitting), we would find a nearly discontinuous switch.

\begin{figure*}
    \centering
    \includegraphics[width=\linewidth]{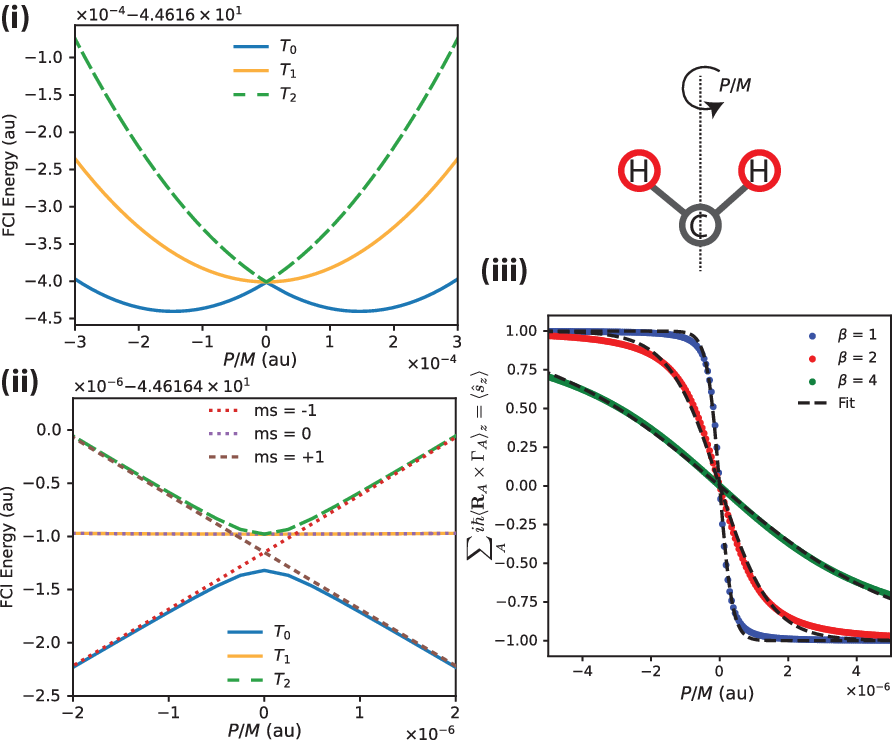}
    \caption{(i) FCI energy contours of the lowest 3 triplets of CH$_2$ in STO-3G basis, including (ii) zoomed in perspective showing the avoided crossing that arises with SOC; we also include spin diabatic energies here. (iii) The expectation value of the cross-product of position with $\hat{\bm \Gamma}$ on the ground state along a normalized unit vibrational mode. Here, we study various artifical strengths ($\beta$) of spin orbit coupling; the dashed lines represent the $\mathrm{tanh}$ fit curves.} 
    \label{fig:ch2}
\end{figure*}

\subsection{Polyatomic CH$_2$ }

Our third  example is the 
polyatomic molecular fragment CH$_2$, the smallest and most infamous ground state triplet in the organic chemistry literature. For this molecule, there are enough degrees of freedom that we cannot scan the entire surface easily. Instead, we begin our analysis by  using analytical gradients on the GHF surface to minimize the geometry and find coordinates $(\bR_0,\bP_0)$,  with minima at C-H bond length 2.04267 au for the bent $C_{2v}$ structure, and minimal velocity for the H atoms $|\bP|= 1.4622\times 10^{-4}$ au (See diagram of the rigid rotation in Fig. \ref{fig:ch2}). At such a stationary point,  the $\{\bf{P}_0\}$ vectors form a rigid molecular rotation (see top right of Fig.\ref{fig:ch2}), and in particular the rigid rotation that minimizes the nuclear moment of inertia.

Following the same analysis as for O$_2$,
in Figure \ref{fig:ch2} (i) and (ii), we plot the FCI curve both over a full range in velocity and a zoomed in range. We also plot the relevant diabatic states in $(ii)$.  Qualitatively, the CH$_2$ data is similar to the O$_2$ data.  However, note that the  CH$_2$ displays a  much smaller spin-orbit coupling as compared with
 O$_2$, which leads to a much smaller $\left| \bP_0\right|$ value and zero field splitting. 

 In order to further explore how the difference in SOC leads to differences between CH$_2$ and O$_2$, we 
 have evaluated a series of artificial calculations on CH$_2$ where we multiply the SOC in the Hamiltonian (Eq. \ref{eq:Hso1}) by a parameter $\beta$. While $\beta = 1$ is physical {\em ab initio} data, if we set $\beta > 1$, we can learn more about symmetry breaking emerges in these degenerate spin state. To that end, in Fig. \ref{fig:ch2}(iii) (see the solid dots), we plot the expectation value of the spin $\hat{\bm{s}}_z$ of the ground  states as function of velocity for different $\beta-$values. As anticipated above, the transition for $m_s = +1$ to $m_s = -1$ becomes smoother for larger $\beta$ values (i.e.  larger SOC values); in other words, if we increase the SOC, we increase the adiabatic gap, and the avoided crossing becomes more adiabatic (as a function of $\bP-$ space).  Numerically, we can find fit all of our numerical data to the form (letting $P$ be the relevant one-dimensional direction in Fig. \ref{fig:ch2}): 
 \begin{eqnarray}
     \sum_A \langle 
      \bm{R_A}\times \hGamma_A\rangle_z (\beta,P) = \mathrm{tanh}(C \beta^2 \frac{P}{M}) \label{eq:tanh}
 \end{eqnarray}
 where $C$ is a constant fixed by the data with $\beta = 1$ ($C= -2.69 \times 10^{-6}$).
A quadratic dependence on $\beta^2$ is consistent with using second-order perturbation theory to evaluate eigenvalues. 
 Two items are now worth noting. First,  the emergence of a $\mathrm{tanh}$ function is reminiscent of the mathematics of  phase transition\cite{Goldenfeld1992-jh}-- even though we are clearly dealing here with an isolated molecule which has no bearing on phase transitions.
Second, it would appear that $\partial^4 E/\partial^2\bm{P}\partial^2\beta$ is the first non-zero mixed derivative between nuclear momentum and spin-orbit coupling, implying that any attempt to study such  symmetry breaking exclusively through response theory along $\bP = 0$ will be difficult and require many derivatives.

 

\begin{figure}
    \centering
    \includegraphics[width=3.25in]{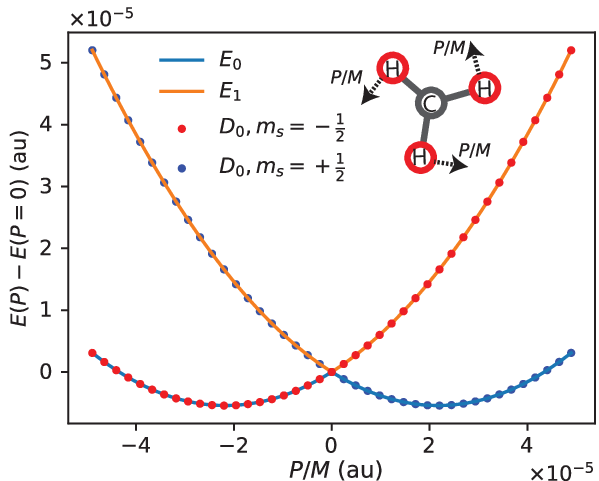}
    \caption{The ground state energies for the $CH_3$ doublet as a function of nuclear $P/M$ for the rigid rotation shown in the in inset. We plot both FCI energies (dots) and GHF energies (line).  Note that, by replacing the Born-Oppenheimer electronic Hamiltonian with a phase-space electronic Hamiltonian, we remove the Kramers' degeneracy when $P \ne 0$. The basis is STO-3G.
}
    \label{fig:CH3}
\end{figure}
 
\subsection{A Doublet: CH$_3^{\bullet}$}

Our fourth and final case is the molecule CH$_3^{\bullet}$,   a radical doublet. 
Following the approach for CH$_2$, we have optimized the phase space energy, where  again $(\bR_0,\bP_0$ indicates that 
the minimizing velocity of the system corresponds to a rigid in-plane rotation of the planar molecule.  In Fig \ref{fig:CH3}, we show the double well that emerges in such a reduced one-dimensional space.  Now, for this problem, it is crucial to emphasize that, because Kramer's degeneracy, we are guaranteed to find exact degeneracy at $v=0$ (which we do). Moreover, if we let the molecule lie in the x-y plane, then $\hat{\bm s}_z$ is completely diagonal in the adiabatic basis. In other words, one cannot diabatize the doublet states of CH$_3$ as one can for the triplet states of CH$_2$. 

Lastly, note that for this problem, FCI is unnecessary; as shown in Fig. \ref{fig:CH3}, a simple GHF ansatz agrees with full CI to a high accuracy  ($> 95 \%$ of the FCI wavefunction); as far as the phenomenology of the PS Hamiltonian goes, the doublet wavefunctions are effectively pure slater determinants.

 \subsection{A classical interpretation of broken symmetry phase space minima}

Before concluding this section, we note that, to zeroth order, some features of the above broken symmetry solutions can be found from a simple, classical model of electron spin. Here, for a diatomic molecule, we model the spin as a small loop of charge rotating locally at an angular frequency ($\omega_s$) around each nucleus.  
Let $(r_1 - R_1)$ and $(r_2- R_2)$ be the distances of the electron ‘spin loop’ from nuclei 1 and 2. 
Now, let us further suppose that the diatomic molecule (and nuclei 1 and 2) is globally rotating with an  angular velocity $\omega_f$ about a fixed origin. The total energy of the system can be represented as:

\begin{equation}
\begin{aligned}
E &= E_1  + E_2 \\
E_1 &= \frac12 m(r_1-R_1)^2(\omega_s-\omega_f)^2   +\frac12 m r_1^2\omega_f^2  + \frac12 M_1 R_1^2 \omega_f^2 \\
E_2 &= \frac12 m(r_2-R_2)^2(\omega_s-\omega_f)^2   +\frac12 m r_2^2\omega_f^2  + \frac12 M_2 R_2^2 \omega_f^2 
\end{aligned}
\end{equation}
For both $E_1$ and $E_2$, the first term is the kinetic energy of the spin particle about the nucleus, the second term is the kinetic energy of the spin arising from the global rotation, and the third term is the kinetic energy of the nucleus arising from the global rotation.

Let us focus on $E_1$ for now and expand the quadratic function (noting that $E_2$ has the exact same functional form).

\begin{equation}
    E_1 = \frac{1}{2} \left( m(r_1-R_1)^2\omega_s^2 - 2m(r_1-R_1)^2\omega_s \cdot \omega_f + m(r_1-R_1)\omega_f^2   + mr_1^2\omega_f^2 + MR_1^2\omega_f^2 \right)\\
\end{equation}

For a spin, we will take the limit that $r_1 \rightarrow R_1$.  However, in doing so, we must recognize  the spin angular momentum,  $\langle \hat{\bm{s}}\rangle = \lim\limits_{r_1 \rightarrow R_1}  I \omega_s = m(r_1-R_1)^2 \omega_s$, need not be zero. That being said, we will ignore the self energy term,  $\frac{1}{2}  m(r_1-R_1)^2\omega_s^2 $, so that:

\begin{equation}
\begin{aligned}
E_1 &=  \frac{1}{2} \left( -2 \hbar \langle\hat{\bm{s}}\rangle \omega_f + M R_1^2 \omega_f^2 + mR_1^2\omega_f^2 \right)\\
\end{aligned}
\end{equation}
 Here, $mR_1^2\omega_f^2$ is effectively a centrifugal term 
 and $2\hbar\langle\hat{\bm{s}}\rangle \omega_f$ is a Coriolis  term, and an identical expression is found for $E_2$.
If we now differentiate $E_1$ with respect fo $\omega_f$ to seek a stationary solution, we find:
\begin{eqnarray}
 \omega_{f,min} = \frac{\hbar \langle \hat{\bm{s}}\rangle  }{2(m+M)R_1^2} \label{eq:classical_w}  
\end{eqnarray}
 Note that we would recover the analogous answer had we worked with $R_2$ instead of $R_1.$
Lastly, 
set $R_1 = R/2$ and our final answer is:
\begin{eqnarray}
    \frac{P_{min}}{M} = \omega_{f,min} R = \frac{\hbar \langle \hat{\bm{s}}\rangle }{2(m+M)R} \label{eq:classical_P/M}   
\end{eqnarray}

Here, $P$ is the absolute value of the momentum in the angular direction in Fig. \ref{fig:H2}.
 Eq. \ref{eq:classical_w} and \ref{eq:classical_P/M} are a simple expression for the  critical $\omega_f$ and $P_{min}$ values  
 as appropriate for each spin eigenstate; these predictions match the functional form of previous classical models based on  perturbation theory as well.\cite{Kayama1967}  For our purposes, in Table \ref{tab:classical}, we show that this model can fairly well (within a factor of 2) recover the stationary points from Figs. \ref{fig:H2} and \ref{fig:O2}-\ref{fig:CH3} above:

\begin{table}[!ht]
    \caption{Table of predicted minimum $P$ (from Eq. \ref{eq:classical_P/M}) versus the empirically determined minima using FCI. Here,  $P$  indicates the momentum in the 1D coordinate from Fig. 
\ref{fig:H2} We use the bond length for which the $m_s = -1$ becomes the ground state in H$_2$. Note that the theoretical model works quite well for diatomics.}
    \centering
    \begin{tabular}{|c|c|c|c|}
    \hline
        ~ & $P_{\mathrm{FCI}}$ & R  & Predicted $P$  \\ \hline
        H$_2$ & 2.3E-05 & 8.5  & 3.2E-05  \\ \hline
        O$_2$ & 2.3E-05 & 2.282  & 1.5E-05  \\ \hline
    \end{tabular}
    \label{tab:classical}
\end{table}

\section{Discussion\label{sec:disc}}

The results above offer a novel view of the coupling between electronic spin and nuclear motion in a fashion that is compatible with angular momentum conservation. In  summary,  we find that a phase space ansatz removes the degeneracy of the ground states for electronic states with doublet or triplet character, and instead associates different spin states with different stationary coordinates in phase space.  At these stationary points, even in the ground state, we find nonzero electronic momentum (linear and angular);  moreover, we are able to ascertain a direction for the electronic spin in three dimensional space. It is important to note that finding such {\em unique} stationary points at $\bP \ne 0$ is possible  {\em only} with an approximate $\hGamma$  in lieu of the true  derivative coupling $\bd$.  After all, if one were to explore the Shenvi phase space Hamiltonian that uses the full derivative coupling, Eq. \ref{eq:Shenvi},  the diagonal element of $\bd$ in Eq. \ref{eq:Shenvi} is gauge-dependent because the BO adiabats are defined only up to a phase.  Moreover, for a problem with an even number of electrons and time reversibility, $\bm{\nabla} \times \bd_{00}^A = 0$ so that one can always choose 
$\bd_{00}^A =0$ (so that $\bP^A_{min} = 0$ is guaranteed); but in systems with multiple degenerate spin states (e.g. a triplet), the meaning of the on-diagonal Berry curvature is limited because the off-diagonal elements in the non-abelian curvature can be large and there is no reason one should necessarily set $\bd_{00}^A =0$. As discussed in Sec. \ref{sec:interpretation}, one of the interesting consequences of choosing an approximate $\hGamma$ is that we can break time-reversibility and find stationary points uniquely without the gauge problems inherent in Eq. \ref{eq:Shenvi}.

Now, given the surprising finding of $\bP_{min} \ne 0$, there are two questions to be asked. First, noting that the key features that allows for such a solution is the presence of a spin-Coriolis term in Eq. \ref{eq:H_PS} that lacks the fine structure constant in Eq. \ref{eq:Hso1}, one can ask: if and when is Eq. \ref{eq:H_PS} valid? Second, given the small size of the minima in Fig. \ref{fig:H2}, and the small barrier size between the two minima ((i) in Fig. \ref{fig:O2}) one can ask: does the existence of these broken symmetry states have any practical consequences. Let us now answer these two questions in turn. 

\subsection{Validity of the Spin Coriolis Effect\label{sub:validity}}

Let us begin with the question of the validity of a  phase space spin-Coriolis force and Eqs. \ref{eq:H_PS} and \ref{eq:Gamma_s} above.  
Physically, a Coriolis force emerges whenever we study dynamics in a moving frame.  Therefore, quite generally, one must ask: when is it more accurate or less accurate to run dynamics in a moving (as opposed to an inertial) frame.

Let us ignore spin for a moment. Recall that the essence of the BO approximation is that, because electrons and nuclei interact strongly (with matrix elements on the order of one atomic unit), it would seem natural to solve for the electronic degrees of freedom in the frame of the nuclei, i.e. we choose orbitals for an electronic basis that depend on nuclear position. Mathematically, this instinct is confirmed by estimating the errors that arise after the BO approximation, namely the Born-Huang correction:
\begin{eqnarray}
    \hat{H}'_{BH} \propto \frac{\hbar^2 \bm{d} \cdot \bP}{M} \label{eq:BornHuang}
\end{eqnarray}
where $\bm{d}$ is the derivative coupling and $M$ is a nuclear mass.
If one is far from a BO state crossing, then Eq. \ref{eq:cond1} above makes clear that $\bm{d}$ can be approximated by $-i \bp/\hbar$ -- which is also on the order of one atomic unit. Therefore, if we make the BO approximation, we expect errors are on the order of $\bP/M$ in atomic units.
It remains only to quantify the size of the $\bP$ element. Of course, the actual size of $\bP$ must depend on the dynamics of interest; nevertheless, according to Born-Huang theory, one can make the simple estimate that (i) if we consider a vibration, then  $\bP \propto (m_e/M)^{-1/4}$; (ii) if we consider a rotation of a molecule, then   $\bP$ is of order unity in atomic units.  (Remember that $m_e$ is one in atomic units.)  The bottom line is that one justifies the BO approximation and solving for the electronic states with frozen nuclei because the expected energy errors are on the order $(m_e/M)^{3/4}$ or $m_e/M$  in atomic units -- whereas we would expect errors are on the order of one atomic unit if we were to solve the Schrodinger equation with a fixed set of electronic orbitals that did not depend on nuclear position.

Now, what about spin? Clearly, if there were no spin-orbit coupling, then it would not make sense to work in a spin basis that rotates with the nuclear frame; one should just fix the spin basis in the lab frame. That being said, however, with enough SOC, it is imperative to solve for the spin degrees of freedom in the molecular frame. {\edit  In the world of molecular rotational spectroscopy, these situations would be determined by comparing  measured spectra to the idealized Hund's cases of angular momentum coupling, where cases all cases except (b) include some degree of coupling between the nuclear rotation and spin (often mediated through intermediate coupling between l and s).\cite{Brown2003} To return to first principals, we}  can find an estimate for when exactly we must allow the spin to rotate with the nuclei by following the same logic as above. Of all of the matrix elements that couple spin together in Eqs. \ref{eq:HSR}-\ref{eq:Hso3}, let us focus on Eq. \ref{eq:Hso1}. If we assume that the  distance between the electron and a nearby nucleus are on the order of one atomic unit (i.e. a Bohr radius) and the electronic momentum ($\left<\hat{\bp}_i\right>$)   and spin ($\left<\hat{\bm s}_i\right>$) are also about one atomic unit, then the magnitude of Eq. \ref{eq:Hso1} is roughly $Z \alpha^2$ where $Z$ is a nuclear charge. In other words, if we were to fix the direction of a spin (independent of nuclear geometry) and then solve the Schrodinger equation, then the errors that would arise are on the order of $Z \alpha^2$.
We may now compare $Z \alpha^2$ vs. $m_e/M$ or  $(m_e/M)^{3/4}$
Their ratio is:
\begin{eqnarray}
    \beta_{vib} &=& \frac{Z \alpha^2}{(m_e/M)^{3/4}}\\
    \beta_{rot} &=& \frac{Z \alpha^2}{(m_e/M)} 
\end{eqnarray}
If we further assume that, for a typical atom, $M \approx 2ZM_H$, we further find:
\begin{eqnarray}
    \beta_{vib} &=& \frac{2^{3/4}Z^{7/4} \alpha^2}{(m_e/M_H)^{3/4}} = 0.025 Z^{7/4} \\
    \beta_{rot} &=& \frac{2Z^2 \alpha^2}{(m_e/M_H)} = 0.20 Z^2\\
\end{eqnarray}
where $m_e/M_H = 1/1837$.

Therefore, at the end of the day, $\beta_{rot} \approx 1$ if $Z = 2.2$
and $\beta_{vib} \approx 1$ if $Z =8.2$.  Thus, for low frequency rotations, solving the Schrodinger equation with a spin that rotates in a body frame should be increasingly more accurate than fixing the spin in a lab frame for systems with atoms larger than a lithium (Z=3, i.e. almost every molecule or material except hydrogen); for vibrations, we expect that mapping the spin to a body frame is reasonable for systems with atoms larger than a fluorine (Z = 9). 
Moreover, for such systems, we can expect that, once we take into account the spin-Coriolis force and Eq. \ref{eq:Gamma_s}, we will find much stronger spin effects than anticipated by spin-orbit coupling alone (because Eq. \ref{eq:Gamma_s} lacks the $\alpha^2$ prefactor). {\edit Lastly, we note for the spectroscopist's perspective, such a simplistic order of magnitude estimates will not differentiate between several of Hund's cases that are also determined by the total strength of the SOC.  That being said,  however, such an estimate would exclude molecules that are a pure case (b) where one cannot justify aligning the spin to the nuclear frame. In practice, it is widely acknowledged that most organic molecules lie on a spectrum between cases (a) and (b), and case (c), where the PS Hamiltonian is most justified; moreover, we anticipate the PS Hamiltonian to be most applicable to heavier atom systems more likely to be found in solids.}

 Obviously, the orders of magnitude that we estimate above will need to be verified by much more accurate models in future work; one should also aim to construct spectroscopic predictions in the future and compare against experiment. That being said, the take-home is clear: For hydrogen, where SOC is tiny, we should {\em not} boost the spin in the body frame; one never runs dynamics on the phase space potential energy surface in Fig. \ref{fig:contour}.  However, for most common molecules and materials (including the molecules in Fig. \ref{fig:O2}-\ref{fig:CH3}) with first row elements,  we probably should solve for spin degrees of freedom adiabatically rather than perturbatively;  for second or lower row atoms, we {\em must} solve for spin degrees of freedom adiabatically for maximum accuracy. Admittedly, the SOC for carbon is small compared to other electronic energy scales (which might tempt the electronic structure theorist to use perturbation theory). However,  such SOC values are not small compared with the nuclear-electronic coupling scales that are the bedrock of BO theory. Moreover, in such a limit, we can also expect very new spin physics to emerge because of the spin-Coriolis force and the resulting symmetry breaking potential energy surfaces described above. 
 
At this point, however, one fundamental and practical question remains regarding the electronic phase space Hamiltonian proposed above for large systems. 
In order to use Eqs. \ref{eq:H_PS} -\ref{eq:Gamma_s} above and go beyond BO theory, one must set the length scale for locality by the parameter $\sigma$ and the localizing function $\zeta$. Now, the parameter  $\sigma$ defines a length scale that separates atoms--which is not hard to guess (roughly an angstrom or a bohr).  That being said, the localization function $\zeta$ is far more interesting and dictates 
how long range is the body frame that dictates  how we orient the spin, and therefore the size and extent of $\Gamma_A''^{(s)}$.
Of course, one could also wonder whether one need choose the same body frame for orienting the electronic orbital momentum and the spin; should there be the equivalent of two different $\zeta$ functions, $\zeta_{AB}^{(l)}$ and  
$\zeta_{AB}^{(s)}$?

Note that we have effectively not  addressed this question above numerically; for molecules with only a few atoms and optimal $\bP$ that are rigid rotations (all the molecules included in this work)), we find no significant localization effects using Eqs. \ref{eq:R_B^0} and \ref{eq:K_B}. {\edit Additionally, it is already known that under such global rotations, an effective magnetic field induced by the nuclear motion can be invoked to reproduce the spin symmetry breaking, equivalent to solving the Born Oppenheimer problem in curvilinear coordinates.\cite{Pickard2002,Ceresoli2002,Stengel2018}} That being said, if we aim to study molecular motion that involves flexible, chaotic nuclear environments, a {\edit locality} cutoff must be made on physical (rather than mathematical) grounds.  After all, if we consider an electron circulating in the vicinity of atom $A$, there is no reason to align the frame of that atom's orbital and spin motion relative to a vector that points from atom $A$ to another atom $B$ located a mile away.  Given that the Coriolis force arises from solving the electronic structure within a rotating nuclear frame, the unavoidable question is: what is the length scale that establishes how large a distance defines such a rotating frame? {\edit Concurrently, experimental techniques have only recently been devised to explicitly observe the coupling between internal motion and spins.\cite{Coudert2025}} Thus, future investigations will inevitably need to study the effects of spin locality in bigger molecules with multiple functional groups, which has much broader implications for chemical reactivity, and specifically {\em ab initio} theory of the CISS effect. 

\subsection{Collectivity, the Size of the Spin-Coriolis Force, and the Einstein de Haas effect}

The questions above about size extensivity lead us directly into the nature of the still not fully understood Einstein-de Haas effect\cite{einstein_dehaas_1915a,einstein_dehaas_1915b}. As a brief reminder, the Einstein de Hass effect is relevant to both micro-\cite{Ganzhorn2016_einsteindehaas,Wallis2006_EdH_in_NiFe} and macroscopic magnetic materials, where under a reversal of a magnetic field, conservation of angular momentum dictates that the material begin to rotate in the lab-frame.
One can ask the obvious question: on what potential surface does the metal slab rotate? Certainly the answer cannot be the Born-Oppenheimer potential energy surface, $E_{BO}(\bm R)$, which is independent of rotational angle, {\edit{a fact which has motivated a great deal of recent physics.\cite{lemesheko:2015:prl,lemesheko:2016:prx,lemesheko:2017:prl}}}.

\begin{figure}
    \centering
    \includegraphics[width=\linewidth]{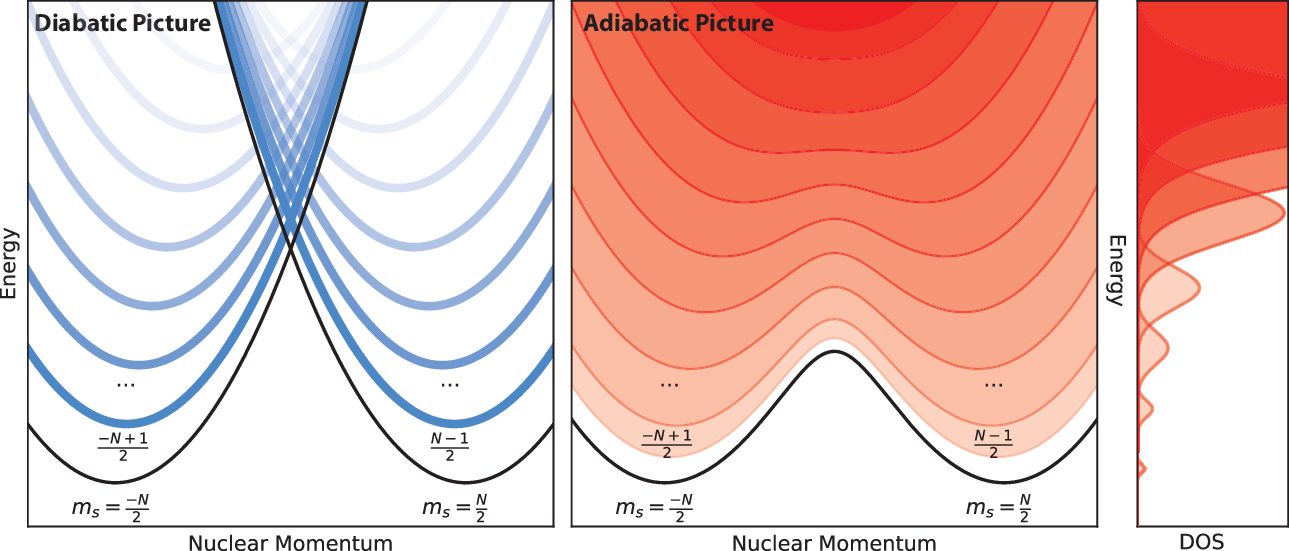}
    \caption{Hypothetical figure showing the diabatic and adiabatic surfaces on which nuclei move during non-adiabatic dynamics that describe the Einstein de Haas effect.}
    \label{fig:EdH}
\end{figure}
That being said, 
 one must wonder if phase space electronic Hamiltonian surfaces are applicable and, given the potential energy pictures in Figs. \ref{fig:O2}-\ref{fig:CH3} above, whether a similar figure can be drawn for the Einstein-de Haas effect. In particular, note from Eq. \ref{eq:Gamma=l+s} above, that we expect that $\langle \hGamma \rangle$  must be macroscopically large for a ferromagnetic with $\langle \hat{\bm{s}}\rangle \ne 0.$
To that end, we hypothesize that, if we were to simulate the potential energy surface for a huge slab of a ferromagnetic field, we would find a result matching Fig. \ref{fig:EdH} with very large double wells in $\bP$.  We submit that, upon a reversal in magnetization direction, one must move from one well to another well along such a phase space surface. For such motion, we are guaranteed to conserve the total momentum, and therefore the nuclear $\bm{P}$ operates as a surrogate order parameter for spin $\bm{\hat{s}}$. Note that, in Fig. \ref{fig:EdH}, we plot not only the ground state but also many excited states; according to Eq. \ref{eq:Gamma=p}, we expect the size of $\left|\bP_{min}\right|$ to be proportional to $\langle \hat{\bm{s}} \rangle$, so that the higher excited states (with less spin polarization) have less displaced double minima.

Interestingly, the proposed picture  Einstein de Haas non-adiabatic dynamics (Fig. \ref{fig:EdH}) is consistent with experiments observed by Tauchert et. al., in which pump driven ultrafast demagnetization of a Ni sheet leads to the creation of phonons carrying angular momentum ($\sim$ 500 fs), causing nuclear displacement in the plane perpendicular to the lost spin.\cite{Tauchert2022_photonEdH} These highly anisotropic phonons then lead to macroscopic rotation of the sample on  a much slower nuclear time scale. Of course, to model such dynamics in the future, we necessarily need to run dynamics along many different potential energy surfaces (with crossings between spin states) and not simply generate potential energy surface. 

Looking forward, there is now a great deal of work to do in order to validate or disprove the picture above for Einstein-de Haas and spin crossing physics more generally. Besides the question of exploring $\zeta$ (see Sec. \ref{sub:validity} above), as a  practical next step, our most immediate concern is to implement the phase space electronic Hamiltonian (Eqs. \ref{eq:H_PS}-\ref{eq:Gamma_s}) within a periodic solid-state electronic structure package. 
Moving to the periodic regime  presents challenges to the founding design principle of a phase space electronic Hamiltonian. For instance, it is not obvious that such a momentum conserving reference frame transformation will be easily implemented within  Born-Von Karman boundaries (e.g. it is known that metals in such systems already violate the acoustic sum rule for vibrations \cite{Dreyer2022}) highlighting that one must treat momentum carefully within periodic boundary conditions. In particular, one must wonder if we will need to modify the angular momentum constraint in Eq. \ref{eq:cond3s} when angular momentum is more difficult to capture in a periodic calculation\cite{BurgosAtencia2024}? On  a philosophical level, these questions play into the broader questions of when molecules become materials, and where does the inherent geometry of space have consequences (and then what are the implications for spin)? 

\section{Conclusions}

In conclusion, we find that, when we solve electronic structure problems with spin orbit coupling in a basis that rotates with the nuclear body frame (orienting both electronic spatial and spin coordinates to that body frame), a large spin-dependent momentum coupling $\hat{\bm\Gamma}_A''^{(s)}$ (Eq. \ref{eq:Gamma_s}) appears that is independent of the fine-structure constant (and yields a large spin-Coriolis force).  One consequence of this spin-dependent term is that a phase space electronic Hamiltonian predicts that the 
degeneracy of electronic spin states are broken.  In particular, for doublets, we find two different minima in $\bP$-space corresponding to two different spin states; for triplets, we also find two minima corresponding to $\ket{\uparrow \uparrow}$ and $\ket{\downarrow \downarrow}$ (the $m_s = 0$ state is usually higher energy).  The implication of this analysis is that, for systems with spin orbit coupling, one should find {\em nonzero electronic orbital and spin angular momentum even in the ground state}. We believe these results should be valid for molecules in the first row of periodic table (that are usually thought to have small spin-orbit coupling). 

The investigations of small molecules presented here  are only the beginning of a long  journey. First, we hypothesize that such a non-perturbative coupling between nuclear movement and spin provides a framework for the \textit{ab initio} simulation of the Einstein de Haas effect. Second, there is currently a great deal of interest in the CISS effect, where most models today cannot explain the strong spin-dependence of electron transfer for systems with relatively weak SOC; to that end, the spin-Coriolis force inherent in  $\hat{\bm\Gamma}_A''^{(s)}$ may well prove very useful.  Third, we wonder whether the framework provided here provides an adequate description of strong correlation between so-called `heavy' electrons and phonons so that we can use phase space electronic structure to study superconducting materials. For instance, might we be able to redraw potential energy the diagrams in Fig. \ref{fig:EdH} for antiferromagnetic and superconducting materials where the two ground state wells are both centered at $P = 0$ and the excited states are spread out over larger nuclear momenta (rather than the inverse)? In such a framework,  might  superconductivity arises from a coherence between two such wells? There is still  a great deal to learn about about phase space electronic Hamiltonians.

\acknowledgments{
This work has been supported by the U.S. Department of Energy, Office of Science, Office of Basic Energy Sciences, under Award no. DE-SC0025393.
}

\bibliography{main}
\end{document}